\RequirePackage{fix-cm}
\documentclass[twocolumn,epjc3]{svjour3}  
\usepackage{aas_macros}
\usepackage{graphicx}
\smartqed  
\journalname{Eur. Phys. J. A}
\begin{document}

\title{
Equation of state and neutrino transfer in supernovae and neutron stars
}


\author{Kohsuke Sumiyoshi\thanksref{e1,addr1}
}

\thankstext{e1}{e-mail: sumi@numazu-ct.ac.jp}


\institute{National Institute of Technology, Numazu College, 
Ooka 3600, Numazu, Shizuoka 410-8501, Japan
\label{addr1}
}

\date{Received: date / Accepted: date}

\maketitle

\begin{abstract}
We overview the progress of the tables of the equation of state for astrophysical simulations and the numerical methods of neutrino transfer.  
Hot and dense matter play essential roles in core-collapse supernovae and neutron stars.  
Equation of state determines the structure of compact objects and their dynamics through its behavior of thermodynamic quantities.  
In addition, neutrinos are trapped in supernova cores and neutron star mergers and frequently interact with matter to crucially affect dynamics in determining the explosion mechanism and the final form of compact objects.  
Therefore, it is essential to implement detailed processes of nuclear and neutrino physics in numerical simulations by having reliable data set of the equation of state and reaction rates.  
We show examples of developments of the equation of state and the neutrino transfer and discuss research directions toward understanding the explosive phenomena by the first principle calculation.  

\end{abstract}
%

%
\section{Introduction}
\label{sec:intro}
Properties of hot and dense matter play essential roles in core-collapse supernovae, which cause gigantically explosive phenomena and leave neutron stars, cosmic rays, and heavy elements.  
Its environment is extreme: the density, temperature, and composition are beyond the experimentally accessible range realized in atomic nuclei \cite{bet90,jan12a,oer17}.  
The matter encounters a wide range of situations during the explosion dynamics, which is different from the static configuration of cold neutron stars.  
It is mandatory to provide the equation of state (EOS) of the hot and dense matter in a suitable form for numerical simulations.
Therefore, theoretical modeling of EOS for supernovae has been a difficult task.  

There have been extensive studies on 
the EOS for nuclear matter (including neutron star matter) and associated efforts to construct the sets of EOS for supernovae \cite{oer17}.  
During this endeavor, it has been important to match the EOS sets with the constraints on nuclear parameters obtained from experiments and observations.  
It is also important to explore the impact of nuclear parameters of the EOS sets in astrophysical simulations.  
The recent progress of nuclear experiments and neutron star observations
has significant influence in this respect \cite{tsa12,lat13,horo14,oze16,li18,li19}.  
There has been also the progress of nuclear many-body theory to provide the EOS for supernovae \cite{tog17,fur17b,sch19a,fur20,dri21,burg21,log21}.  
These advances call for cautious examinations of the EOS for supernovae and their impact on supernova simulations.  

Meanwhile, numerical simulations of core-collapse supernovae have remarkable progress in revealing the explosion mechanism \cite{jan12b,bur13,kot13,jan16} thanks to increasing resources of high-performance computing.  
The neutrino heating mechanism is commonly believed to proceed together with non-spherical hydrodynamics to have a successful explosion.  
There are, though, still issues on robustness and magnitude of explosion related with uncertainties due to microphysics and numerical methods.  
One of the difficult problems is the coupling of neutrino and matter, which determines the energy transfer to help the explosion.  

Recently there is progress by the numerical simulations which solve the neutrino transfer without approximations by solving directly the Boltzmann equation \cite{sum12,nag14,nag18}.  
The uncertainty of neutrino heating is closely related to approximate numerical methods to solve the neutrino transfer, which describes the neutrino reactions and propagation.  
The exact approach of neutrino transfer is helpful to pin down the unknown factors in the neutrino heating in multi-dimensional phenomena.  
Accordingly, it is now urgent to examine the impact of nuclear physics: supernova EOS through the neutrino interaction.  

In this article, we overview the progress of supernova  EOS over decades and discuss the direction of researches in recent and coming years.  
We describe the efforts of microscopic many-body theories and systematic approaches of the supernova EOS to examine the influence on the dynamics.  
We explain the neutrino heating mechanism in the studies of core-collapse supernovae with an emphasis on neutrino transfer.  
Full numerical simulations by solving the Boltzmann equation are underway for core-collapse supernovae and neutron star merger.  
We stress that the exact treatment of neutrino transfer is contributing to reveal EOS effects for the successful explosion.  
We discuss the necessity of extension of data entries for neutrino processes on CompOSE\footnote{\texttt{https://compose.obspm.fr}} \cite{typ21} for the full treatment of microphysics in the advanced technology of astrophysical simulations.  

\section{History of supernova EOS}
\label{sec:historyEOS}
We briefly overview the history of EOS sets for numerical simulations of core-collapse supernovae.  
Ever since the first numerical simulations \cite{col66}, there have been extensive efforts to implement the essence of EOS with multi-composition in a wide range of conditions.  
To investigate the influence of hot and dense matter, systematic studies utilizing analytic formulae have been made in the 1980s \cite{bar85,tak88}.  
Dependence on the nuclear parameters such as the incompressibility and adiabatic index has been studied to explore the outcome of explosions.  
It has been demonstrated that the softness of EOS is important to gain large gravitational energy by compression of central core for successful explosions.  
The basic idea of the description of hot and dense matter in multi-phase of nucleons and nuclei has been established around this era \cite{lam81}.  

Construction of the set of EOS suitable for supernova simulations, however, is a painstaking task due to the necessity of completeness to cover the wide range of density, temperature, and composition.  
The data table of EOS by Wolff and Hillebrandt is the first such product aimed at numerical simulations \cite{hil84}.  
The set of thermodynamical quantities of the hot and dense matter is evaluated by the sophisticated Skyrme Hartree-Fock calculations for finite nuclei in a Wigner-Seitz cell.  
This artwork is the beginning of the EOS data, which sets the early standard of supernova EOS for later studies.  However, this EOS table has not been publicly available.  

The first publicly available set of supernova EOS has been published in a form of computer code by \cite{lat91} (LS-EOS).  
The EOS is constructed by the compressible liquid-drop model for nuclei using the functional form of energy of nuclear matter based on the Skyrme interaction.  
It has three options for the value of incompressibility to cover the uncertainty of nuclear parameters.  
This EOS set has been popularly used for some time as a unique choice.  

As the second publicly available set, the data table of supernova EOS has been published by Shen et al. \cite{she98a,she98b} (Shen-EOS) acting later as a counterpart of LS-EOS.  
The EOS is constructed by the relativistic mean-field (RMF) theory with the Thomas-Fermi approximation.  
The usage of the RMF theory is motivated by the description of the saturation properties of nuclear matter in the relativistic Br\"{u}ckner-Hartree-Fock theory \cite{bro90,sug94}.  
It draws interesting attention through comparisons with non-relativistic many-body approaches as employed in LS-EOS.  
The data table covers the wide range of density, proton fraction, and temperature including the limiting value of zero temperature and zero proton fraction, which are often requested for numerical simulations.  
These two sets of EOS (LS-EOS and Shen-EOS) have been routinely used in astrophysical simulations over the years.  

Various improvements of supernova EOS have been made afterward based on the wide usage of the supernova EOS \cite{oer17}.  
The effective interaction of the RMF theory has been revised with the knowledge of experiments and observations.  
In \cite{ste13}, new interactions are constructed to match the neutron star observations \cite{ste10}, for example.  
The density-dependence of the symmetry energy to reproduce the nuclear collective modes \cite{rut05} is adopted to construct the EOS table \cite{hem12}.  
The density-dependent coupling constants are also adopted to improve the properties of matter and nuclei in \cite{gshe10,typ10,fis14,ban14}.  
The mixture of various nuclei has been taken into account to describe the thermodynamical equilibrium instead of a representative nucleus \cite{hem10,fur11,bli11,gshe11,hem12,ste13,fur13a}.  
A possible mixture of hyperons and occurrence of quark-hadron phase transition is implemented as extensions of the EOS \cite{ish08,nak08a,sag09,fis11,oer12,ban14}.  
There are sets of supernova EOS based on microscopic many-body approaches \cite{tog17,fur17b,sch19a,fur20,dri21,burg21,log21}.  
These developments of the EOS sets are summarized on the web page of the activity, CompOSE, and the data and information of the EOS sets are available in a convenient manner.  

We briefly summarize the extension of the Shen EOS as an example of the continuous advance of supernova EOS.  
The Shen EOS is constructed to provide the numerical data table, being different from LS-EOS, by covering the wide range of environment with completeness and smoothness of data \cite{she98a,she98b}.  
The form of the RMF theory is chosen based on the success of the relativistic many-body frameworks \cite{sug94}.  
Its interaction is fixed by the nuclear masses and radii and has been kept in later developments.  
Therefore, it is possible to solely explore the influence of the extensions.  

Based on the original form of Shen-EOS, the RMF theory extended with strangeness is used to provide the EOS tables with a mixture of hyperons \cite{ish08}.  
The phase transition from the hadron phase in the RMF theory to the quark gas phase is described by utilizing the MIT Bag model for the construction of the extended EOS tables \cite{nak08a}.  
The original Shen EOS table in 1998 are revised by extending the density range with regular density grids \cite{she11}.  
These extensions are used to study the core collapse of massive stars leading to the black hole formation \cite{sum06,sum07,nak08b,sum09}.  
The mixture of nuclei under the nuclear statistical equilibrium is described in recent products of the EOS table \cite{fur11,fur17a} instead of the assumption of single nuclear species in the original Shen EOS.  
The description of the mixture of nuclei is important to describe the electron captures during the gravitational collapse, the shock propagation, and neutrino heating \cite{hix03,sum08b,sul16,fur17c,nag19a,fis20}.  

\section{Developments of EOS}
\label{sec:directionEOS}
In the era of having a variety of sets of supernova EOS, it is necessary to consider new developments of EOS sets with motivations to complement and strengthen the available sets so far.  
Here, we would like to discuss two lines of research: microscopic approaches and systematic variations to explore the influence of EOS.  

It is indispensable to construct the supernova EOS in microscopic many-body theories starting from the nuclear interaction toward the goal of EOS determination.  
It is profitable to examine the characteristics of supernova EOS in microscopic approaches since most of the EOS sets so far have been constructed based on the effective interactions in the mean-field frameworks.  
While the construction of the complete EOS for supernovae is the best, it is also vital to clarify the influence of nuclear physics systematically.  
From the side of numerical simulations in astrophysics, it is essential to explore effects due to uncertainty piece by piece, in our case, nuclear parameters, for example.  
Therefore, it is important to provide the EOS sets in a systematic form within effective approaches so that one can explore its influence (see also \cite{sch17,nak19}, for example).  

\subsection{Microscopic approaches}
\label{sec:microscopicEOS}
We describe the recent developments of the supernova EOS by the microscopic approaches as examples.  
The data table of supernova EOS is constructed by the cluster variational method (VM) for nuclear many-body system \cite{tog17}.  
The uniform nuclear matter is obtained starting from the nuclear potentials supplemented with three-body force, which fulfills the saturation properties of nuclear matter \cite{tog13}.  
The Thomas-Fermi calculation is performed to obtain the configuration of non-uniform matter with extension to finite temperature.  
It is further extended to the EOS set with multi-composition of nuclei under the nuclear statistical equilibrium \cite{fur17b}.  
These EOS sets are representatives of the supernova EOS which cover the wide range of density, temperature, and composition based on the non-relativistic many-body theories.  
It is worth mentioning that there are developments of EOS sets based on the non-relativistic frameworks, including the Br\"{u}ckner Hartree-Fock theory and chiral effective field theory \cite{sch19a,dri21,burg21,log21}.  

Recently a counterpart of the EOS set obtained by the relativistic many-body theory has been constructed \cite{fur20}.   The data table of supernova EOS (DBHF-EOS) is obtained based on the results of the Dirac Br\"{u}ckner Hartree-Fock (DBHF) theory, which reproduces the saturation properties of nuclear matter starting from the nucleon-nucleon interaction \cite{kat13}.  
The interaction energies and relativistic nuclear potentials are utilized by a parametrized form to obtain the configurations of non-uniform matter in an extended liquid drop model at finite temperature with consideration of the mixture of nuclei.  

\begin{figure*}[ht!]
\centering
\includegraphics[width=0.45\textwidth]{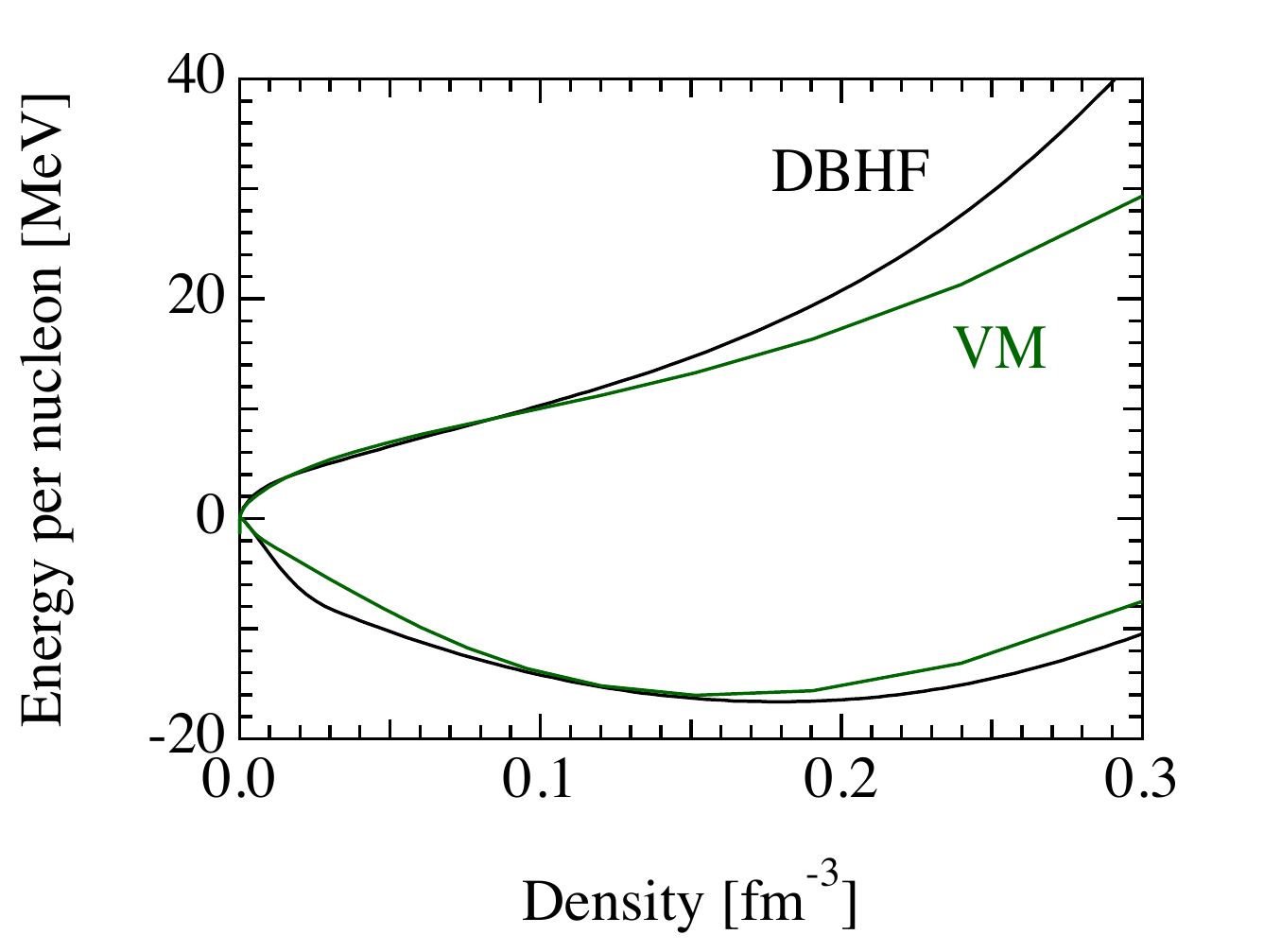}
\includegraphics[width=0.45\textwidth]{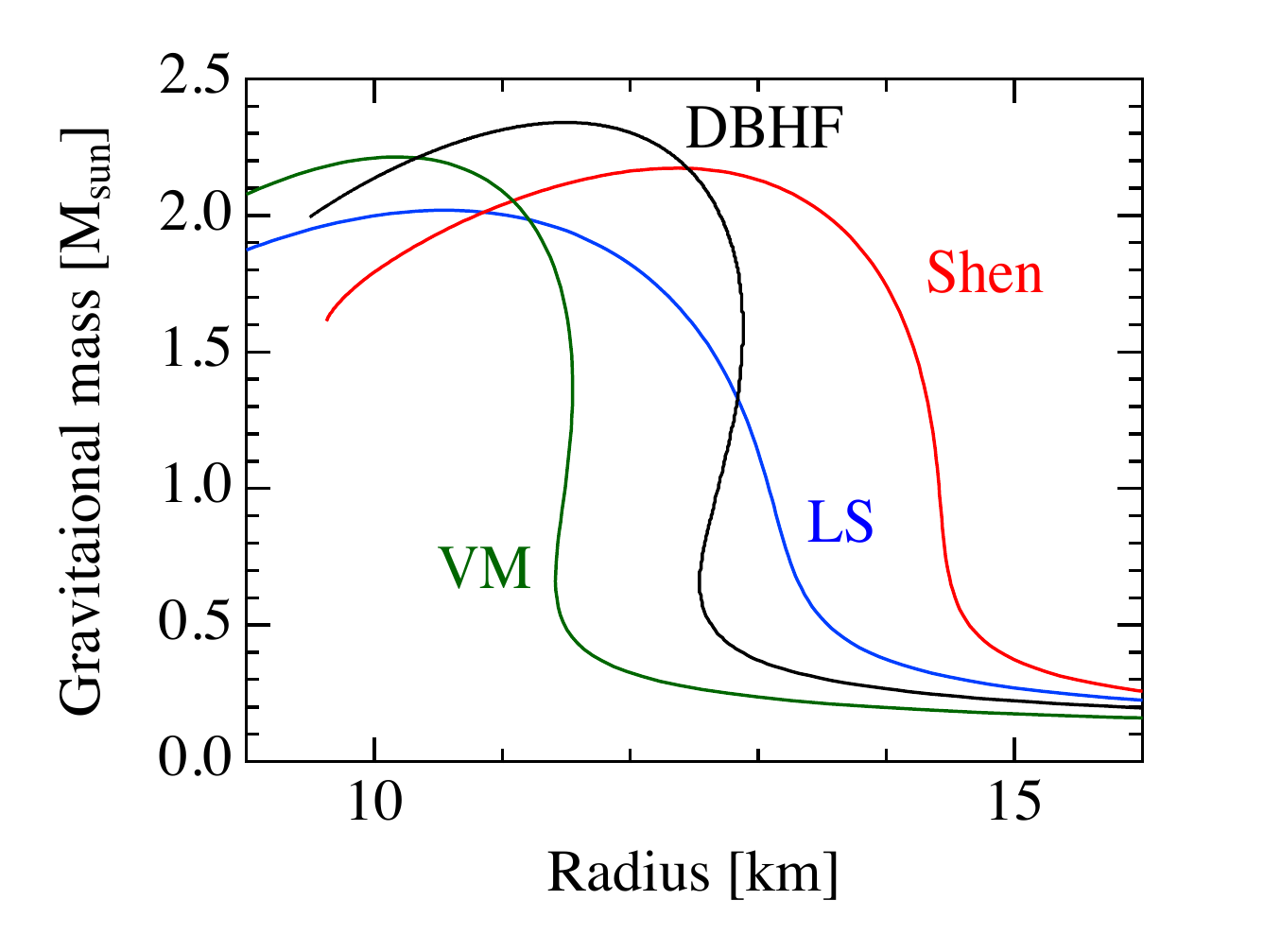}
\caption{Energy per nucleon of symmetric nuclear matter and neutron matter for VM-EOS and DBHF-EOS is shown as a function of the number density of nucleons (left).  Gravitational masses of cold neutron stars as a function of radius for VM-, DBHF-, LS-, and Shen-EOS (right).  
\label{fig:EOS_NS}}
\end{figure*}

The sets of EOS by VM and DBHF reflect different characteristics between non-relativistic and relativistic many-body theories.  
We note that they adopt different nuclear interactions, which also leads to different characteristics.  
Figure \ref{fig:EOS_NS} displays the properties of the two EOS sets \cite{tog13,kat13}.  
Results of symmetric nuclear matter have saturation properties in a successful manner (left panel).  
Energy curves of neutron matter in the two methods are almost the same at low densities.  
In contrast, the energy of neutron matter in DBHF increases faster than that of VM at high densities.  
This increasing feature of symmetry energy is common in relativistic many-body frameworks and is seen in relativistic mean-field theories.  
The relativistic frameworks tend to provide stiff EOS in comparison with the non-relativistic ones.  
The properties of cold neutron stars constructed by the EOS sets reflect such differences in stiffness (right panel).  
The neutron star radii by VM are smaller than those by DBHF.  
These cover the range of the recent observational constraints on neutron star radii \cite{abb18,mcmil19}.  
It is interesting to see that the classic sets of non-relativistic and relativistic EOS (LS and Shen) have similar differences but with larger values.  

The two sets using VM and DBHF energies exhibit interesting differences in the hot and dense matter in supernova cores.  
Besides the different stiffness of uniform matter at high densities, the differences in the composition of nuclei are found in the comparison of two EOS sets obtained in the same framework using VM and DBHF inputs \cite{fur20,fur20b}.  
The DBHF-EOS tends to provide large mass fractions for medium mass nuclei than the VM-EOS.  
This is related to larger values of the saturation density and symmetry energy, which affect the mass of nuclei, in DBHF than those in VM.  
The difference in the composition may affect the gravitational collapse through electron capture processes \cite{hix03,nag19a} and other neutrino-related processes such as neutrino trapping and heating.  

\begin{figure*}[htb!]
\centering
\includegraphics[width=0.45\textwidth]{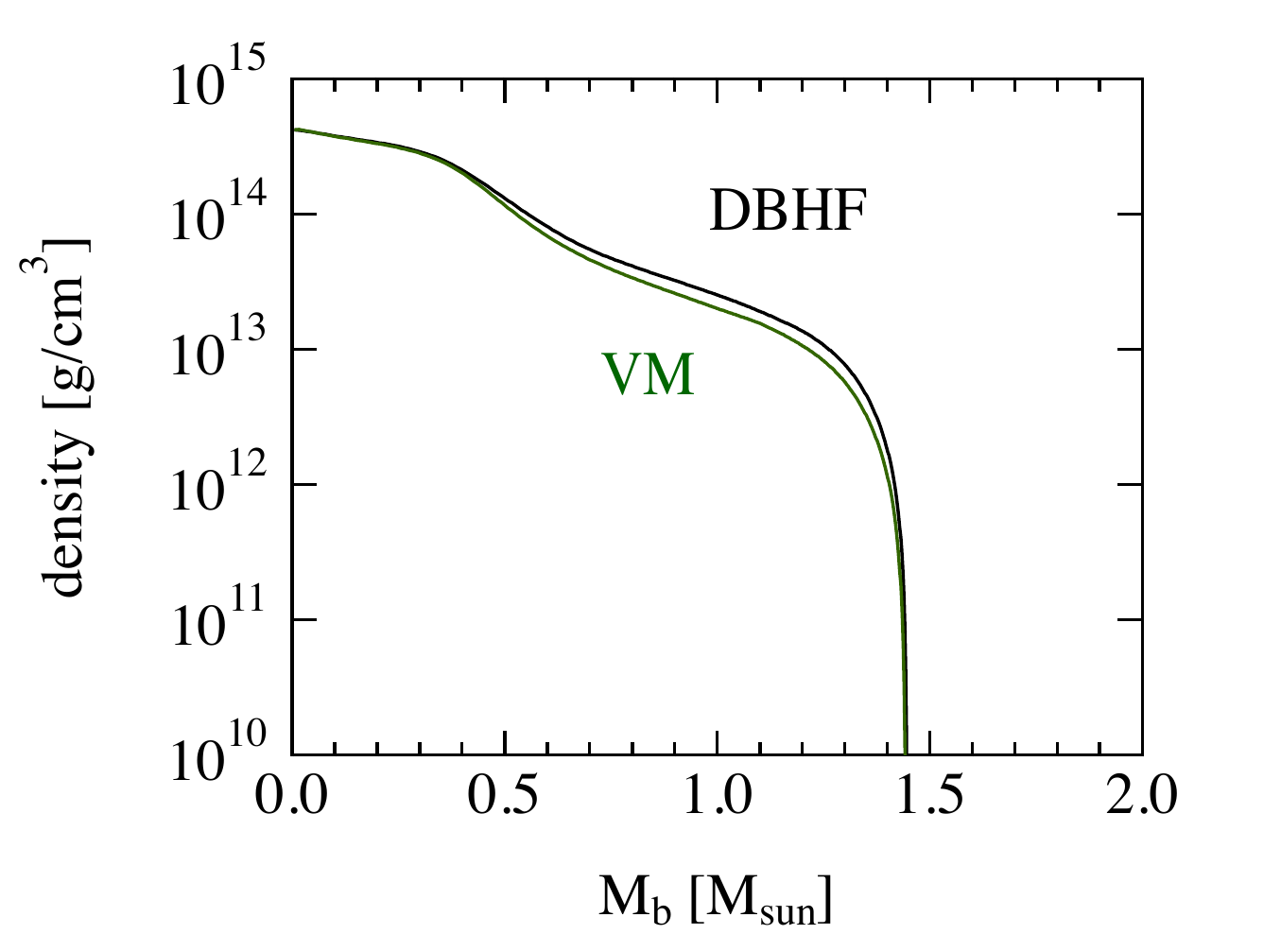}
\includegraphics[width=0.45\textwidth]{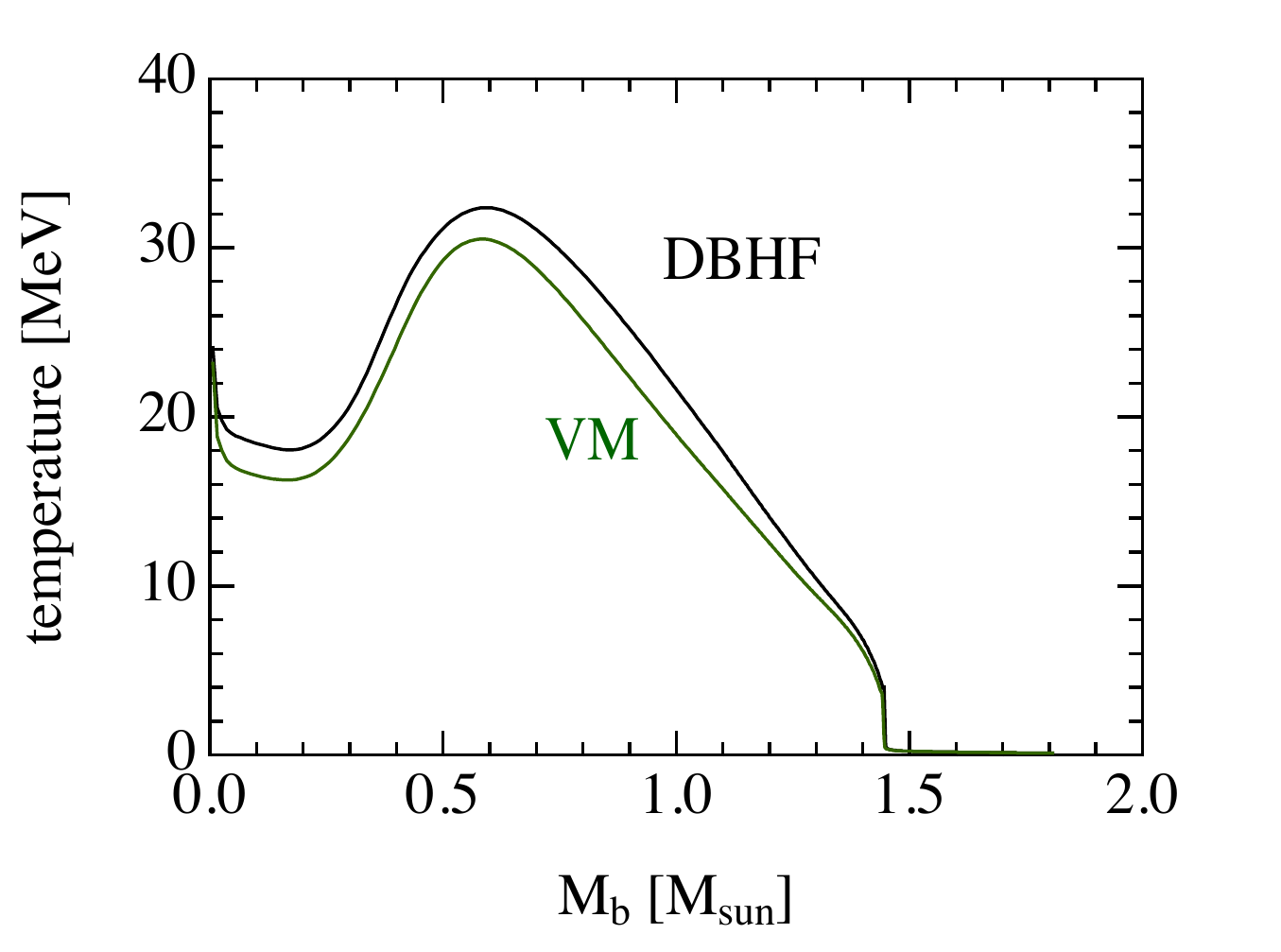}
\caption{Profiles of density (left) and temperature (right) in supernova cores at 200 ms after the bounce of a 15M$_{\odot}$ star are shown by green and black lines for the simulation with VM- and DBHF-EOS, respectively.  
\label{fig:Radhyd_DBHF}}
\end{figure*}

The EOS table by VM (VM-EOS) \cite{tog17} has been utilized in numerical simulations of core-collapse supernovae, black hole formation and proto-neutron star cooling \cite{tog14,nak18,nak21}.  
The softness of the VM-EOS leads to the higher density at the core bounce as compared to the case of Shen-EOS.  
However, it does not provide the explosion of a massive star of 15M$_{\odot}$ \cite{woo95} under the spherical symmetry (see a successful case of a 9.6M$_{\odot}$ star in \cite{nak21}).  
The duration of neutrino signal in the black hole forming collapse of a 40M$_{\odot}$ star \cite{woo95} with VM-EOS is shorter than the case of Shen-EOS but longer than the case of LS-EOS.  
The cooling of proto-neutron star with VM-EOS is slower than the cases of Shen-EOS since the VM-EOS leads to high densities due to the softness and provides a different composition of nuclei \cite{nak18}.  

We show in Fig. \ref{fig:Radhyd_DBHF} comparison of supernova cores in the numerical simulation of the gravitational collapse of a massive star using the VM-EOS \cite{fur17b} and DBHF-EOS \cite{fur20}.  The numerical simulations are performed by solving general relativistic neutrino-radiation hydrodynamics under spherical symmetry \cite{sum05} (see also \S \ref{sec:transferNu}).  
We found the central densities are similar at 200 ms after the bounce and the temperature in the case of DBHF-EOS is slightly higher than the case of VM-EOS.  
Detailed analysis will be reported elsewhere.  
The VM-EOS \cite{fur17b} has been applied to numerical simulations of core-collapse supernovae also in 2D (under axial symmetry) and 3D \cite{nag19a,nag19b,nag19c,iwa20}.  
Numerical results of core-collapse and bounce for an 11.2M$_\odot$ star reveals a rapid expansion of shock wave with an early kick of a proto-neutron star \cite{nag19c}.  
Numerical simulations using DBHF-EOS are underway to make comparisons and will be reported elsewhere.  

\subsection{Systematic approaches}
\label{sec:systematicEOS}
Effective theories and energy functionals motivated by the microscopic nuclear many-body theories are suitable to make systematic studies on the parameters of nuclear matter \cite{swe94,sch17}.  
We describe here an effort to explore the uncertainty in the symmetry energy based on the Shen-EOS obtained by the relativistic mean-field theory.  

The RMF theory starts from the effective Lagrangian density which is composed of nucleon and meson degrees of freedom, 
\begin{eqnarray}
\label{eq:LRMF}
\mathcal{L}_{\rm{RMF}} & = & \sum_{i=p,n}\bar{\psi}_i
\left[ i\gamma_{\mu}\partial^{\mu}-\left(M+g_{\sigma}\sigma\right) 
\right. \nonumber \\
&& \left. -\gamma_{\mu} \left(g_{\omega}\omega^{\mu} +\frac{g_{\rho}}{2}
\tau_a\rho^{a\mu}\right)\right]\psi_i   \nonumber \\
&& +\frac{1}{2}\partial_{\mu}\sigma\partial^{\mu}\sigma -\frac{1}{2}%
m^2_{\sigma}\sigma^2-\frac{1}{3}g_{2}\sigma^{3} -\frac{1}{4}g_{3}\sigma^{4}
\nonumber \\
&& -\frac{1}{4}W_{\mu\nu}W^{\mu\nu} +\frac{1}{2}m^2_{\omega}\omega_{\mu}%
\omega^{\mu} +\frac{1}{4}c_{3}\left(\omega_{\mu}\omega^{\mu}\right)^2  
\nonumber \\
&& -\frac{1}{4}R^a_{\mu\nu}R^{a\mu\nu} +\frac{1}{2}m^2_{\rho}\rho^a_{\mu}%
\rho^{a\mu} 
\nonumber \\
&& +\Lambda_{\rm{v}} \left(g_{\omega}^2
\omega_{\mu}\omega^{\mu}\right)
\left(g_{\rho}^2\rho^a_{\mu}\rho^{a\mu}\right),
\end{eqnarray}
where the standard notation is used.  
The meson-nucleon coupling constants ($g_{\sigma}$, $g_{\omega}$, $g_{\rho}$), meson self-coupling constants ($g_{2}$, $g_{3}$, $c_{3}$), and $\omega$-$\rho$ coupling constant ($\Lambda_{\rm{v}}$) as well as the meson masses ($m_{\sigma}$, $m_{\omega}$, $m_{\rho}$) are the parameter of the theory (See \cite{she11,she20} for details).  

The effective Lagrangian density is originally based on the standard meson-nucleon couplings for the isoscalar scalar and vector mesons ($\sigma$, $\omega$) as well as the isovector vector meson ($\rho$) \cite{ser86,gam90}.  
It has been extended by adding non-linear terms to take into account density-dependence of interactions as the researches of EOS and nuclei progress \cite{bog77,sug94,fat10,bao14a}.  
In the construction of the Shen-EOS, it is extended with non-linear terms of the isoscalar scalar and vector mesons, but without the density-dependent term of the isovector vector meson, on which we will discuss further.  
In the mean-field approximations, the meson fields are treated as classical fields by replacing the field operators with their expectation values.  Through the derivation of the Euler-Lagrange equations, the set of equations for the mean fields of mesons and the Dirac equation for nucleons under the influence of the meson fields is solved in a self-consistent manner \cite{sum94,sum95a}.  
The calculation of the energy momentum tensor provides the energy density and pressure for uniform matter given by 
\begin{eqnarray}
\label{eq:ERMF}
\varepsilon &=& \displaystyle{\sum_{i=p,n} \frac{1}{\pi^2}
  \int_0^{\infty} dk\,k^2\,
  \sqrt{k^2+{M^{\ast}}^2}
  \left( f_{i+}^{k}+f_{i-}^{k}\right)
   } \nonumber\\
 & &
  +\frac{1}{2}m_{\sigma}^2\sigma^2+\frac{1}{3}g_{2}\sigma^{3}
  +\frac{1}{4}g_{3}\sigma^{4} \nonumber\\
 & &
  +\frac{1}{2}m_{\omega}^2\omega^2+\frac{3}{4}c_{3}\omega^{4}
  +\frac{1}{2}m_{\rho}^2\rho^2 \nonumber\\
 & &
  +3 \Lambda_{\rm{v}}\left(g^2_{\omega}\omega^2\right)
     \left(g^2_{\rho}\rho^2\right),
\end{eqnarray}
and
\begin{eqnarray}
\label{eq:PRMF}
 P &=& \displaystyle{\sum_{i=p,n} \frac{1}{3\pi^2}
   \int_0^{\infty} dk\,k^2\,
   \frac{k^2}{\sqrt{k^2+{M^{\ast}}^2}}
   \left( f_{i+}^{k}+f_{i-}^{k}\right)
    } \nonumber\\
 & &
  -\frac{1}{2}m_{\sigma}^2\sigma^2-\frac{1}{3}g_{2}\sigma^{3}
  -\frac{1}{4}g_{3}\sigma^{4} \nonumber\\
 & &
  +\frac{1}{2}m_{\omega}^2\omega^2+\frac{1}{4}c_{3}\omega^{4}
  +\frac{1}{2}m_{\rho}^2\rho^2 \nonumber\\
 & &
  +\Lambda_{\rm{v}}\left(g^2_{\omega}\omega^2\right)
   \left(g^2_{\rho}\rho^2\right).
\end{eqnarray}
Here $M^{\ast}=M+g_{\sigma}\sigma$ is the effective nucleon mass.  
$f_{i+}^{k}$ and $f_{i-}^{k}$ ($i=p,n$) are the occupation
probabilities of nucleon and antinucleon at momentum $k$.  

Since the Shen-EOS has been popularly used in astrophysical simulations, it is meaningful to examine the influence due to its possible uncertainty.  
The effective interaction TM1 of the RMF theory for Shen-EOS has been determined by the least-squares fitting by the calculation of the structure of finite nuclei \cite{sug94}.  
The nuclear mass and radii of nuclei in the wide mass range are used for the fitting to determine the coupling constants of the RMF theory \cite{sug94}.  
It is worth noting that the density-dependent term (the last term of Eq. \ref{eq:LRMF}) is not included for TM1.  
The resulting EOS is categorized in stiff EOSs having the large symmetry energy, which is commonly seen in relativistic many-body theories.  
Note that the determination was made with the limited knowledge of unstable nuclei at that time.  
The bulk properties of uniform matter are listed in Table \ref{tab:EOS}.  
The TM1 interaction has been used continuously in the extensions of the EOS tables to keep the baseline model for comparisons in astrophysical simulations.  
\begin{table}[htbp]
\caption{Properties of uniform matter of the TM1 and TM1e EOS models.  The incompressibility, $K$, symmetry energy, $E_{sym}$, and its slope parameter, $L$, are listed for two models.  }
\centering
\begin{tabular}{lccc}
\hline
     & $K$ [MeV]  & $E_{sym}$ [MeV] & $L$ [MeV] \\
\hline
TM1  & 281 & 36.89 & 110.8 \\
TM1e & 281 & 31.38 & 40    \\
\hline
\end{tabular}
\label{tab:EOS}
\end{table}

Recent experiments and observations provide a better understanding of nuclear properties and motivate us to investigate the character of symmetry energy of Shen-EOS \cite{tsa12,lat13,horo14,oze16,li18,li19}.   
We explore modifications of the symmetry energy of EOS in the RMF theory by adding a density-dependent isovector term (the last term of Eq. \ref{eq:LRMF}) \cite{sum19,hu20}.  
The rapid increase of the symmetry energy can be suppressed by increasing the strength of $\omega$-$\rho$ coupling ($\Lambda_{\rm{v}}$).  
Since the addition is only through the isovector contribution, the properties of the symmetric nuclear matter remain exactly the same.  
By fixing the symmetry energy at a slightly lower density than the saturation density, it is possible to explore different behavior of the density dependence of symmetry energy keeping good properties of stable nuclei \cite{bao14a,bao14b,bao15}.  
Among the variations of the slope parameter, $L$ of the symmetry energy, we choose a case of $L=40$~MeV (TM1e) for the construction of the revised version of Shen-EOS.  
The symmetry energy for TM1e becomes smaller than that for TM1 (Table \ref{tab:EOS}).  

The value of $L=40$~MeV for TM1e is within the current constraints on the symmetry energies while the value $L=110$~MeV for TM1 is claimed too large 
\cite{tsa12,lat13,li18,li19}.  
In fact, the TM1 does not fulfill the criteria in the extensive survey \cite{dut14}.  
However, a recent analysis based on the PREX-2 experiments for the neutron skin thickness derives the range of large value of $L$, which favors TM1 \cite{ree21}, although the extracted values are still controversial \cite{pie21} (See also discussion on neutron star radii below).  

\begin{figure*}[htb!]
\centering
\includegraphics[width=0.45\textwidth]{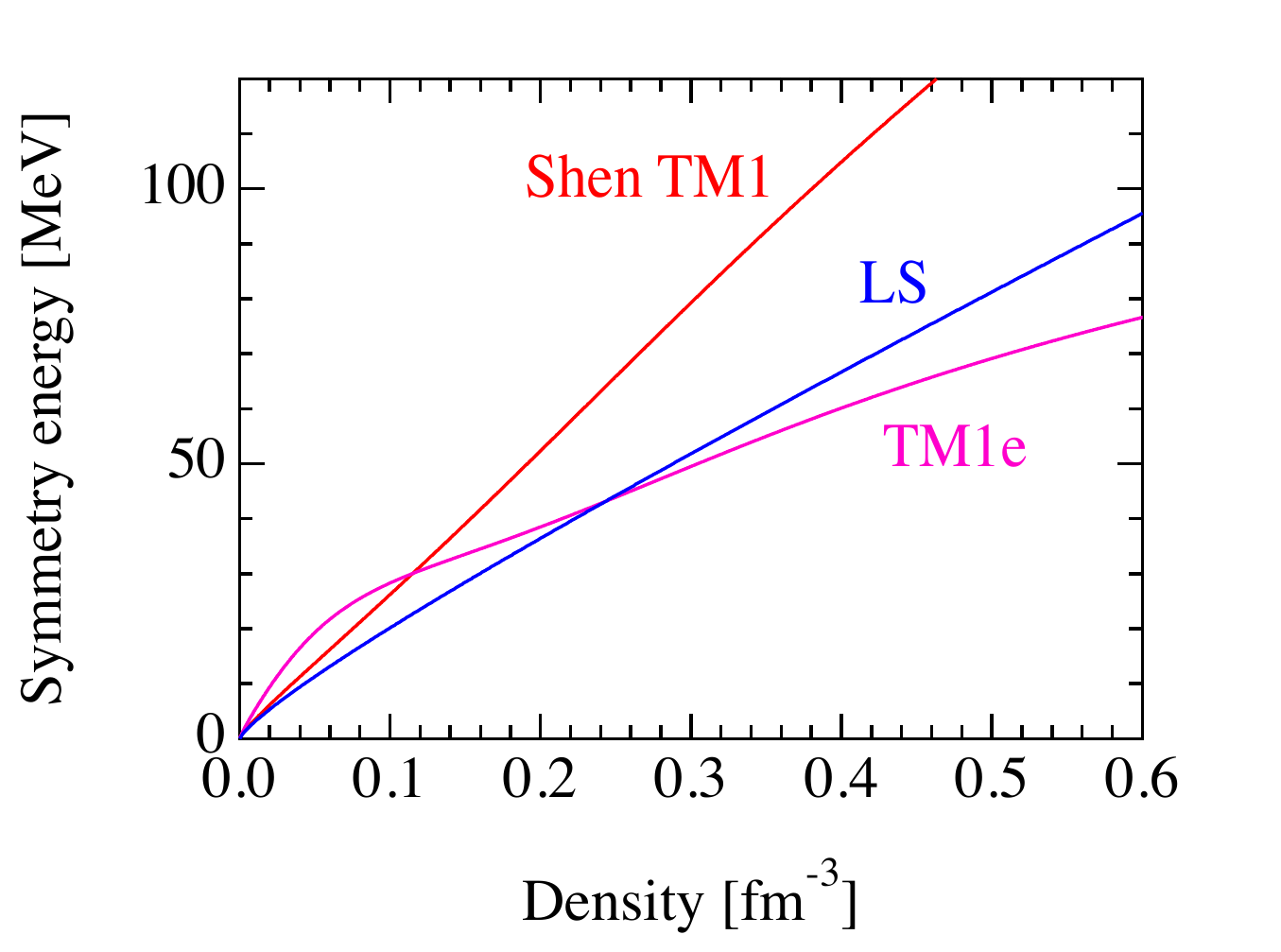}
\includegraphics[width=0.45\textwidth]{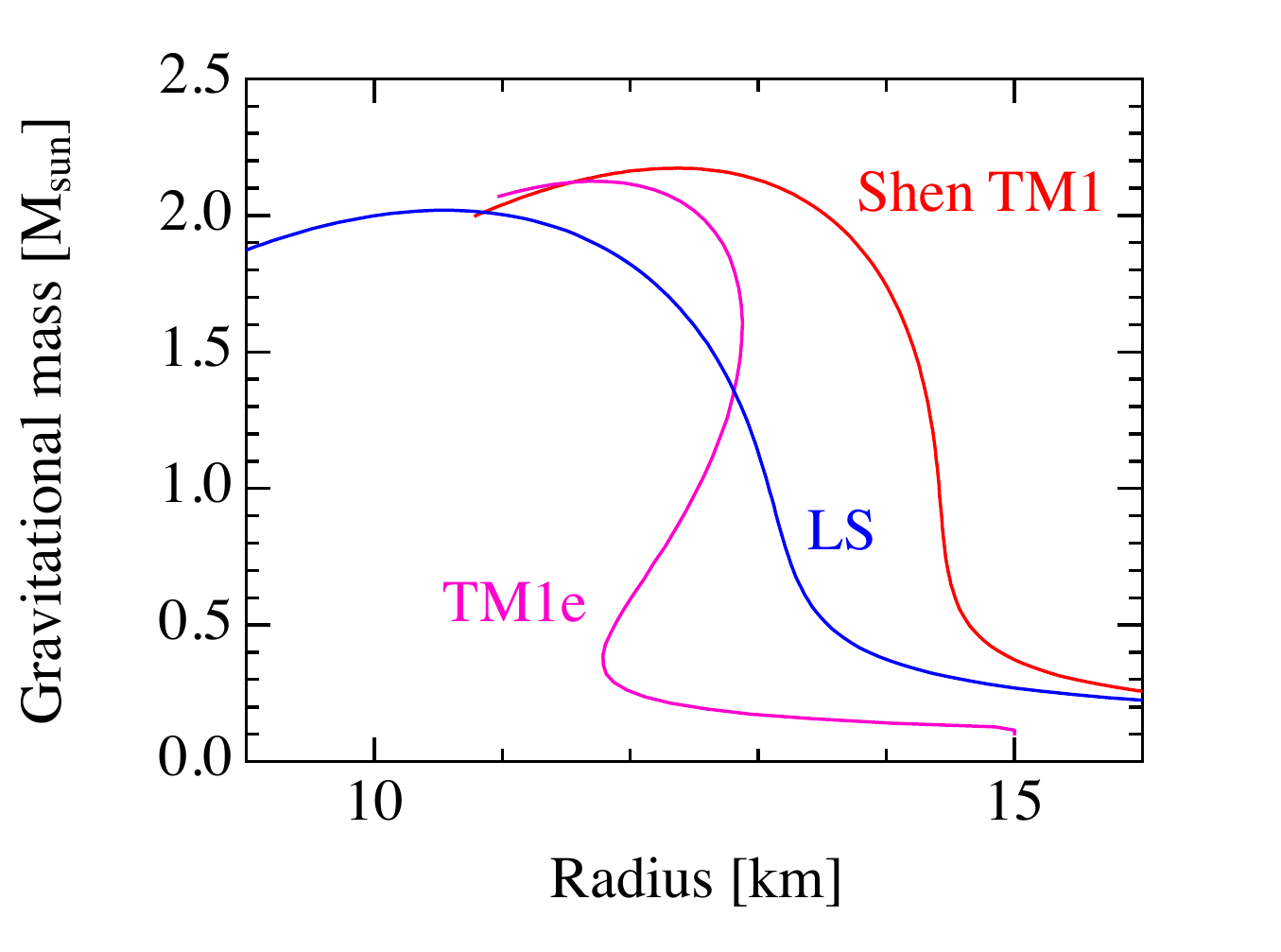}
\caption{Symmetry energy for Shen TM1e, TM1, and LS-EOS is shown as a function of the number density of nucleons (left).  Gravitational masses of cold neutron stars as a function of radius for Shen TM1e, TM1, and LS-EOS (right).  
\label{fig:TM1e_NS}}
\end{figure*}

We show in Fig. \ref{fig:TM1e_NS} the nuclear symmetry energy and the mass-radius relation for neutron stars for TM1 and TM1e \cite{sum19,hu20}.  
The increase of symmetry energy for TM1e is slow at high densities as compared with that for TM1, which provides a rapid linear increase (left panel).  
The radius of cold neutron stars for TM1e is smaller than that for TM1 while the maximum masses for TM1e and TM1 are both beyond 2.0M$_\odot$ (right panel).  
It is worth mentioning that the neutron star radii for TM1e are well within the observational constraints while those for TM1 are too large due to the large symmetry energy.  
It is also interesting to see that the analyses of the recent NICER observations provide rather large radii for a massive neutron star \cite{mcmil21,ril21}.  

The data table of the revised Shen EOS with TM1e has been constructed in the same framework and format as Shen EOS with TM1 and is publicly available \cite{she20}.  
The two sets of EOS with TM1 and TM1e can be used to explore the influence of symmetry energy on the dynamics of core-collapse supernovae and the associated emission of neutrinos.  

\begin{figure*}[htb!]
\centering
\includegraphics[width=0.45\textwidth]{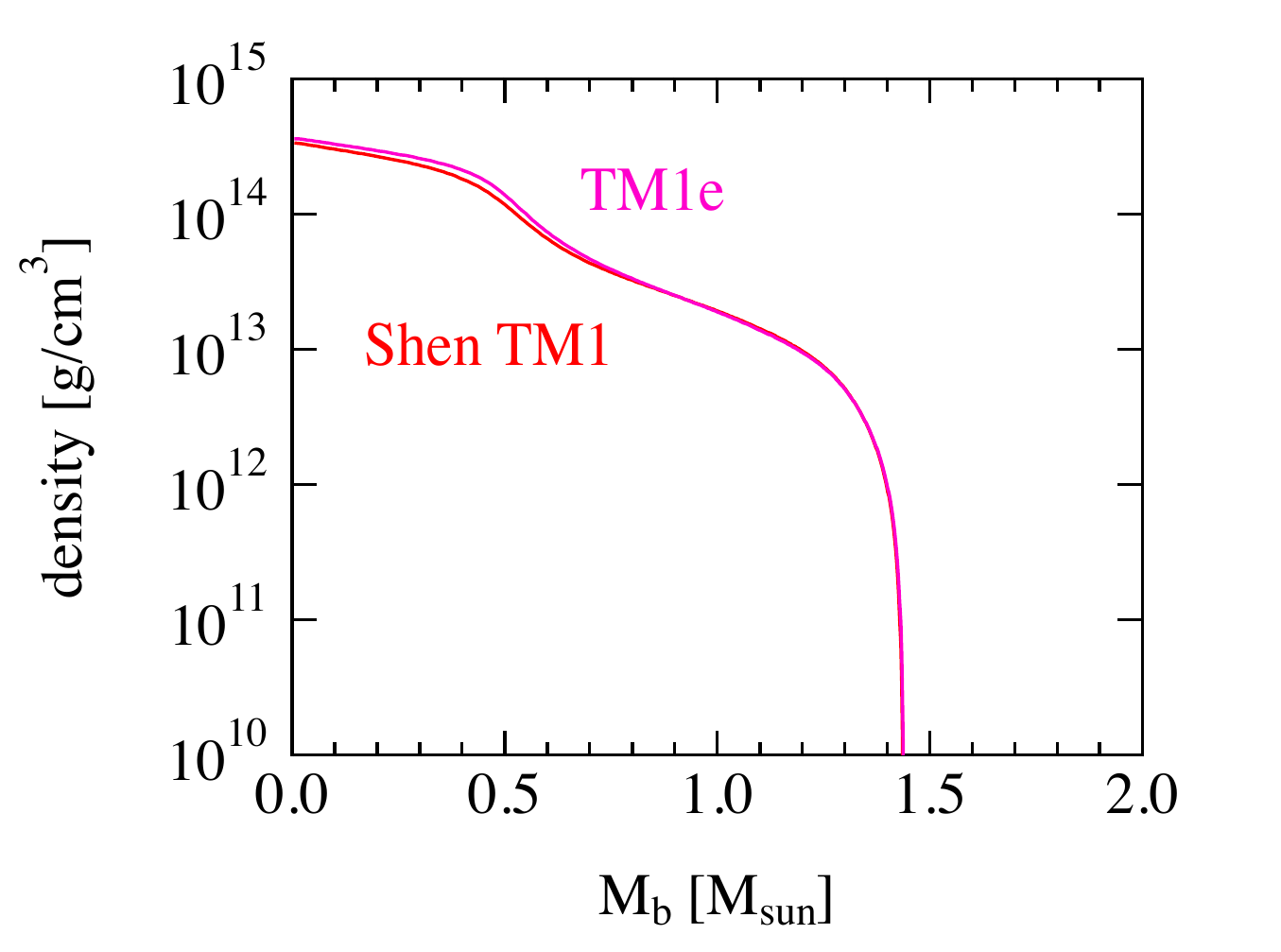}
\includegraphics[width=0.45\textwidth]{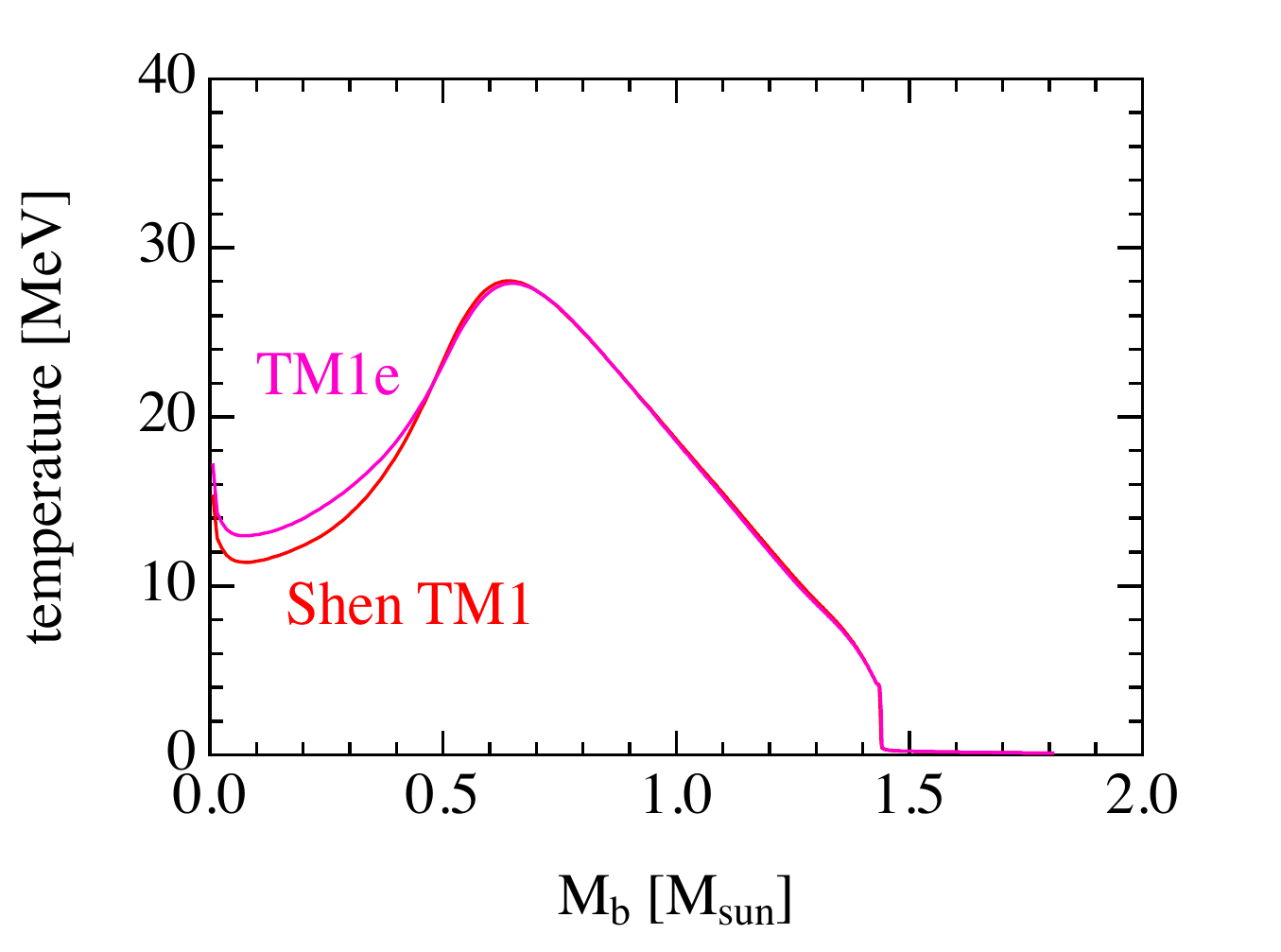}
\caption{Profiles of density (left) and temperature (right) in supernova cores at 200 ms after the bounce of the 15M$_{\odot}$ star are shown by red and magenta lines for the simulation using the sets of EOS with TM1 and TM1e, respectively.  
\label{fig:Radhyd_TM1e}}
\end{figure*}
It is found that profiles of the central core are rather similar around the core bounce.  
Differences appear more clearly in the late phase of proto-neutron star cooling \cite{sum19}.  
We show in Fig. \ref{fig:Radhyd_TM1e} profiles of supernova cores in the numerical simulations of the gravitational collapse of a massive star using the sets of EOS with TM1 \cite{she11} and TM1e \cite{she20}.  
The density and temperature in the case of TM1e are slightly higher than those in the case of TM1.  
However, the two sets of EOS do not provide a large enough difference right after the core bounce.  

This is because the difference between TM1e and TM1 appears only in a neutron-rich environment.  
The proton fraction is not so small ($\sim$0.3) and the density is just above the nuclear saturation density in the central part of the supernova core at the core bounce and, hence, the difference due to the EOSs is not so drastic.  
As the matter becomes neutron-rich and the central density becomes high 
during the cooling of nascent proto-neutron stars, the differences by the symmetry energy become evident \cite{sum19}.  
The resulting emission of neutrinos from proto-neutron stars in TM1e and TM1 shows quantitative differences in the time evolution of energies and luminosities reflecting the density and temperature profiles.  
Note that the EOS with TM1e provides a shorter duration of neutrino bursts than the EOS with TM1 in the black hole forming collapse of massive stars \cite{sum19}.  

\section{Neutrinos in supernovae and neutron stars}
\label{sec:neutrinoSNandNS}

\subsection{Neutrino mechanism in core-collapse supernovae}
\label{sec:neutrinoSN}

Neutrinos and their interaction in hot and dense matter play an essential role in core-collapse supernovae.  
They affect the dynamics of gravitational collapse, bounce, and shock propagation through weak interactions.  
The neutrinos are produced in the collapsing massive star, trapped inside a central core during the collapse, and are emitted as supernova neutrinos.  
Through these stages, they contribute to the explosion dynamics as agents to exchange leptons and energy with matter and affect the propagation of the shock wave.  
The shock wave is initially launched by the core bounce, but its propagation inevitably stalls due to obstacles of free-falling matter from the outer layers of the massive star.  
A small portion of the emitted neutrinos is absorbed by the material and effectively heats the matter behind the stalled shock wave.  
This mechanism of neutrino heating contributes to assisting the revival of outward propagation of the shock wave \cite{bet85,bet90}.  

The neutrino heating mechanism is one of the key issues in the current understanding of the mechanism of the supernova explosion.  
Numerical studies have been extensively made to clarify the role of neutrinos through descriptions of hydrodynamics with neutrino reactions and propagation.  
In order to investigate the neutrinos in the supernova core, it is required to provide the properties of hot and dense matter to determine the rates of neutrino reactions with targets (nucleons, nuclei, and leptons).  
In this regard, providing neutrino reaction rates together with the EOS set is of great importance.  

\subsection{Neutrino transfer in supernovae}
\label{sec:transferNu}
Neutrino transfer, which describes the propagation and reactions of neutrinos in the matter, is one of the difficult problems as in the case of radiation transfer in astrophysics and engineering.  
The propagation and reactions with matter of neutrinos are in principle governed by the Boltzmann equation, 
\begin{equation}
\frac{1}{c} \frac{\partial f_{\nu}}{\partial t} + \vec{n} \cdot \nabla f_{\nu}
= \left[ \frac{1}{c} \frac{\partial f_{\nu}}{\partial t},  \right]_{coll}
\label{eqn:Dynamics_Boltzmann_3DXYZ}
\end{equation}
where $f_{\nu}$ is the neutrino distribution in six dimensions (three space and three neutrino momentum space) and \vec{n} is the unit vector of neutrino propagation.  The left hand side describes the number change of neutrinos through propagation and the right hand side is the collision term for the number change of neutrinos through reactions with matter.  
It is difficult to directly solve the Boltzmann equation and one usually takes approximate methods.  
When one integrates the Boltzmann equation (Eq. \ref{eqn:Dynamics_Boltzmann_3DXYZ}) by neutrino angle with a factor of the neutrino energy, it leads to the continuity equation for the neutrino energy density, 
\begin{equation}
\frac{\partial {\cal E}_{\nu}}{\partial t} + \nabla \cdot \vec{{\cal F}}_{\nu}
= -Q_{\nu}
\label{eqn:Dynamics_Boltzmann_3DEnergy_Conservation}
\end{equation}
where ${\cal E}_{\nu}$, $\vec{{\cal F}}_{\nu}$, and $Q_{\nu}$ are the energy density, energy flux of neutrinos, and energy change via neutrino reactions, respectively.  
If the diffusion approximation can be applied, for example, one can express the neutrino energy flux by the gradient of energy density using the relation, $\vec{{\cal F}}_{\nu}= -D \nabla {\cal E}_{\nu}$ so that it is reduced to the diffusion equation.  
When the neutrinos propagate almost freely, one can adopt the relation for free-streaming instead.  

In core collapse supernovae, the situation of neutrino transfer is non-trivial and one cannot simply justify to take such approximations.  
One has to describe the wide range of conditions from the diffusion deep inside to the free-streaming in outer layers.  
It is a delicate problem because the neutrino heating occurs in the intermediate regime, which is not diffusive or free-streaming.  
Approximate methods may lead to overestimation or underestimation of the size of neutrino heating due to the sensitivity to neutrino distributions in angle, energy, and space.  

It is, hence, ideal to describe the neutrino transfer in a precise manner by avoiding approximations.  
However, the calculation of neutrino transfer requires high computational costs and it has been difficult to perform the full computation of neutrino transfer (see below).  
The first principle calculations of neutrino transfer became possible under the spherical symmetry in the 2000s \cite{ram00,lie01,sum05}.  
In the spherical simulations, it has been established that the shock wave stalls after the launch by core bounce and there is no revival of the shock wave (See, however, \cite{fis11}, for example).  
No explosion can be obtained for ordinary models of massive stars (see also \cite{hue10,mel15,nak21,mor21} for successful cases with low mass stars).  

It is commonly considered that breaking the spherical symmetry is mandatory to have explosions following the failure of spherical simulations \cite{jan12b,bur13,kot13,jan16}.  
The neutrino heating works in combination with hydrodynamical instabilities such as convection seen in numerical simulations in 2D (axially symmetric) and 3D space.  
Multi-dimensional simulations have been extensively performed to find the general tendency of neutrino heating in a longer time scale, which is realized by a non-spherical motion of matter with deformation of shock wave and convection, than that in the direct free-fall of spherical layers.  

However, there are remaining issues in order to fully understand the explosion mechanism.  
On the one hand, it is mandatory to investigate the approximate treatments currently made in the numerical treatment of general relativity and neutrino transfer.  
The validation of approximations can be made by the first principle calculation of neutrino-radiation hydrodynamics in general relativity (including the general relativistic version of Eq. \ref{eqn:Dynamics_Boltzmann_3DXYZ}, for example).  
On the other hand, it is necessary to explore the influence of neutrino and nuclear physics: equation of state and weak reaction rates.  
These physics processes enter (as source terms, for example) in the equations of neutrino-radiation hydrodynamics and remain as uncertainties even in the highly sophisticated numerical treatments.  
One has to proceed with the investigation on these aspects step by step.  
We argue for the growing importance of microphysics having progress of numerical simulations toward the first principle calculations as we describe below.  

Numerical studies of supernovae in 2D/3D space by treating the neutrino transfer without approximations recently became possible with the numerical codes to solve the Boltzmann equation in the full dimensions \cite{sum12,nag14,nag18}.  
Solving the Boltzmann equation is a formidable task since one has to describe the time evolution of the neutrino distribution functions in 3 space and 3 momentum space dimensions.  
Evaluation of the source term of the Boltzmann equation is complicated due to energy and angle dependence, which gives a wide range of time scales and makes the equation numerically stiff, with non-linear terms for pair processes.  
The numerical studies by solving the Boltzmann equation have been made, thanks to high-performance computing resources in these years, and applied to examine the neutrino transfer in 2D/3D supernova cores and to inspect approximate methods such as diffusion, ray-by-ray, and moment closure \cite{sum12,sher17}.  

The 2D/3D calculations of neutrino-radiation hydrodynamics for the dynamics of core-collapse supernovae have been made by coupling the solver of the Boltzmann equation, 2D/3D hydrodynamics and Newtonian gravity \cite{nag14,nag17,nag19b}.  
This is an important step to remove the uncertainty in neutrino transfer, which remained in the previous works with approximate methods, toward the full first principle calculations in general relativity.  
Core-collapse simulations in 2D axial symmetry starting from initial models for the central core of massive stars having 11.2M$_\odot$ and 15M$_\odot$ have been performed to examine the explosion dynamics based on the neutrino heating mechanism and hydrodynamical instabilities \cite{nag18,nag19c,har19,har20}.  
Recently the numerical studies by the Boltzmann equation have been extended to 3D core-collapse supernovae \cite{iwa20} and neutrino transfer in general relativity \cite{aka20}.  
In these movements, the numerical treatment is elaborated to the one for the first principle calculation where the neutrino and nuclear physics can be examined in detail.  

\subsection{EOS influence}
\label{sec:influenceEOS}
It is now the situation to examine the nuclear physics uncertainties by removing the approximation of neutrino transfer in the supernova simulations.  
We focus here on the influence of the EOS on the supernova mechanism under the detailed treatment of neutrino transfer.  
Effects of the neutrino reactions should be also studied along with the EOS influence (see \cite{kot18,nag19a,fis20,betran20}, for example) by considering medium effects and compositional differences.  

Studies of EOS dependence in 2D so far suggested the tendency having successful explosions with the soft case with LS EOS as compared with the stiff cases with Shen-EOS and Wolff-Hillebrandt EOS \cite{jan12a,suw13,fis14}.  
Influence of the nuclear parameters such as the effective mass has been under discussion recently \cite{yas18,sch19b}.  
We explored the influence of EOS on the explosion in 2D simulations by solving the Boltzmann equation with the two sets of EOS: LS-EOS and Furusawa-EOS \cite{fur17b}, which is an extended version of Shen-EOS with the treatment of multi-composition for nuclei.  
It has been shown that the shock wave largely expands in an asymmetric manner with LS-EOS and is closer to an explosion as compared with the case of Furusawa-EOS, which leaves a merely stalled shock \cite{nag18,har20}.  

Closer examinations on the difference of shock dynamics due to the two EOS sets reveals that an early occurrence of convection in LS EOS right after the core bounce triggers large deformation of the shock wave, leading to further shock expansion \cite{har20}.  
It is found that different compositions of matter in the two EOS sets lead to affect the degree of convective activity.  
This is interestingly different from a simple argument by the difference in softness and stiffness.  
Right after the core bounce, there is a stage called prompt convection, which is triggered by the negative gradient in the entropy profile.  
Due to the passage of the shock wave, dissociation of nuclei into nucleons proceeds in dense matter and leads to loss of energy.  
This leads to the decrease of entropy at the post-shock region and makes a negative gradient, which implies hot material resides below cool material in the gravity, as a trigger of the prompt convection.  
The degree of dissociation of nuclei is different for the two EOSs having a different composition in the hot and dense matter in the falling material onto the shock wave.  
It is found that more massive nuclei with a larger mass fraction of nuclei appear in LS-EOS than in Furusawa-EOS.  
The case of LS-EOS leads to a larger loss at the shock passage and a larger negative gradient, which causes earlier and stronger activity of convection.  
Therefore, it suggests that the composition of matter is also important in addition to the softness and stiffness.  
Further studies to explore the influence of different treatments of composition are underway.  
It is also necessary to consistently provide both the equation of state and neutrino reaction rates \cite{pin12,fis14,nag19a}.  

\subsection{Neutrino transfer in neutron star merger}
\label{sec:mergerNS}

Neutrino reactions are involved in the merger of compact objects and influence the evolution of their remnants and peripheral material.  
In the event of binary neutron star merger, GW170817 \cite{Abbott2017a}, the signal of the gravitational wave with associated electromagnetic waves has been detected and detailed information on the central object and the nucleosynthesis of heavy elements has been explored \cite{Abbott2017b,Radice2018a,Shibata2019}.  

Neutrinos are produced in the hot and dense matter during the merger and temporarily trapped inside the massive proto-neutron star and surrounding torus.  
The trapped neutrinos are emitted gradually from the central remnant and contribute to the transfer of energy and modification of composition in the material.  
Emission of neutrinos in the central remnant act as a source of cooling and deleptonization, which drives the thermal evolution toward a black hole or cold neutron star.  
The absorption of neutrinos in the surrounding material causes heating, which enhances the matter ejection and a possible formation of jet phenomena.  
It also contributes to a change in the balance of protons and neutrons and may have an impact on neutron-richness for the r-process nucleosynthesis.  

Neutrinos in the hot and dense matter are, therefore, important in the study of binary neutron star mergers as in the case of core-collapse supernovae.  
Neutrino processes have been implemented in the studies of numerical simulations of mergers to explore their influence.  
Various approaches of neutrino transfer have been taken with different levels of approximations to perform numerical simulations for the evolution of binary and merger with mass ejection (see \cite{end20,sum2021} for further references).  
It is stressed that the numerical computation of mergers in general relativity is expensive and the full treatment of neutrino transfer is a grand challenge.  
Neutrino transfer in the merger has different significance from the one in core-collapse supernovae, where the threshold for explosion or failure sensitively depends on the neutrino transfer.  

Recently properties of neutrino transfer in the remnant of binary neutron star merger have been studied by directly solving the multi-dimensional Boltzmann equation \cite{sum2021}.  
The neutrino distributions in a deformed massive neutron star with a surrounding torus have been obtained in multi-dimensions (space, neutrino angle, and energy).  
It has been shown that the neutrino fluxes from the deformed remnant are highly asymmetric and the neutrinosphere has a largely deformed shape extended to the torus.  
The information of calculated neutrino reaction rates in hot and dense matter shows the neutrino reactions are important in and above the massive neutron star.  
The full information of neutrino distributions and neutrino reaction rates is valuable for validation of approximate methods and developments of numerical prescriptions for neutrino processes in numerical relativity.  
Hence, it is becoming more important to investigate the equation of state with neutrino reaction rates in neutron star merger having the detailed treatment of neutrino transfer.  

\section{Discussion and summary}

As the numerical studies of supernovae and neutron stars are progressing with the elaborated methods on high-performance computing resources, providing data in nuclear and neutrino physics becomes more vital to reveal the explosive phenomena in a comprehensive manner.  
To pin down questions in the complicated explosion mechanism, it is crucial to provide the nuclear data with reduced uncertainties.  
In addition, it is necessary to reveal the role of the equation of state and neutrinos by examining systematic trends in numerical studies.  
Hence, coordinated efforts to provide the data tables of the equation of state with detailed information are indispensable.  
The activity on the CompOSE, for example, is helpful to share the information of the EOS tables in a common format with useful tools.  
It would be further productive to implement the information of neutrino reaction rates, such as NuLib\footnote{\texttt{https://github.com/evanoconnor/NuLib}}, together with the equation of state since the reaction rates depend on the properties of hot and dense matter.  

It is important to study further the equation of state and neutrino processes in hot and dense matter inside the binary neutron star merger.  
Its extreme environment, having low proton fraction (neutron-rich) and high temperature, is different from the one in supernovae and cold neutron stars.  
In the case of neutron star merger with the neutron-rich environment, the exotic phase may appear at high densities more likely than the supernova case.  
Providing the EOS table with neutrino reaction rates for neutron star merger is also getting more important.   
Communities of numerical relativity are demanding to have the EOS at a very low density and temperature range to describe ambient material around the remnants by smoothly connecting with the ordinary EOS tables.  
It is also important to consider the systematic coverage of the equation of state to search the gravitational wave and to study ejected matter for multi-messenger observations.  

Since the first principle type calculation is just around the corner in the coming decade, it is essential to focus on microphysics to clarify the essential point of the explosion mechanism and neutron star mergers.  
It is becoming more important to clarify the uncertainties of nuclear and neutrino physics in numerical studies.  
We are entering the next phase of providing nuclear data by having reliable data set with the ability to make systematic variations.  

\section{Acknowledgement}
\label{sec:ack}

The author is grateful to S. Furusawa and H. Togashi for their careful reading and helpful comments.  
This work is supported by Grant-in-Aid for Scientific Research (17H06357, 17H06365, 19K03837, 20H01905) from MEXT, Japan.  
For providing high-performance computing resources KEK, JLDG, RCNP of Osaka University, YITP of Kyoto University, and the University of Tokyo are acknowledged.

This work is done under the Particle, Nuclear, and Astro Physics Simulation Program (Nos. 2020-004, 2021-004) of the Institute of Particle and Nuclear Studies, High Energy Accelerator Research Organization (KEK) and supported by "Program for Promoting Researches on the Supercomputer Fugaku" 
(Toward a unified view of the universe: from large scale structures to planets) of MEXT.

\bibliographystyle{spphys}
\bibliography{sumi}{}


%
%

\end{document}